\documentclass[times]{asjcauth}

\usepackage{moreverb}

\usepackage[pdftex,colorlinks,bookmarksopen,bookmarksnumbered,citecolor=red,urlcolor=red]{hyperref}

\usepackage[]{graphicx}
\usepackage{amsmath,amsfonts,epic,eepic,epsfig,subfigure,amssymb}

\usepackage{graphics} 
\usepackage{mathptmx} 
\usepackage{times} 
\usepackage{url}
\usepackage{algorithm,algorithmicx}
\usepackage{algpseudocode}
\usepackage{lineno}
\usepackage{bbm}
\usepackage{cite}

\def\volumeyear{2018}

\begin{document}

\runninghead{P.~R.~Barbosa et al.: Controlled Tracking in Urban Terrain: Closing the Loop}


\title{CONTROLLED TRACKING IN URBAN TERRAIN: CLOSING THE LOOP}

\author{Patricia~R.~Barbosa, Yugandhar~Sarkale, Edwin~K.~P.~Chong, Yun~Li, Sofia~Suvorova and Bill~Moran}
\address{Patricia R. Barbosa is with Rockwell Collins, Richardson, TX 75082, USA (e-mail: pbarbosa@ieee.org).\\\hspace*{0.3cm}Yugandhar Sarkale (corresponding author, e-mail: Yugandhar.Sarkale@colostate.edu), and Edwin K. P. Chong (e-mail: Edwin.Chong@colostate.edu) are with the Department of Electrical and Computer Engineering, Colorado State University, Fort Collins, CO 80523-1373, USA.\\\hspace*{0.3cm}Yun Li is with Hillcrest Labs, Rockville, MD 20850, USA.\\\hspace*{0.3cm}Sofia Suvorova (e-mail: sofia.suvorova@unimelb.edu.au), and Bill Moran (e-mail: wmoran@unimelb.edu.au) are with the Department of Electrical and Electronic Engineering, University of Melbourne, Parkville, Vic 3010, Australia.}

\received{April 30, 2018.}

\acks{This research was supported in part by DARPA under contract FA8750-05-2-0285.\\\hspace*{0.3cm}The material in this paper was presented in part at the SPIE Conference on Signal and Data Processing of Small Targets, Orlando, FL, March 17-20, 2008.}

\begin{abstract}
We investigate the challenging problem of integrating detection, signal processing, target tracking, and adaptive waveform scheduling with \emph{lookahead} in urban terrain. We propose a closed-loop active sensing system to address this problem by exploiting three distinct levels of diversity: (1) spatial diversity through the use of coordinated multistatic radars; (2) waveform diversity by adaptively scheduling the transmitted waveform; and (3) motion model diversity by using a bank of parallel filters matched to different motion models. Specifically, at every radar scan, the waveform that yields the minimum trace of the one-step-ahead error covariance matrix is transmitted; the received signal goes through a matched-filter, and curve fitting is used to extract range and range-rate measurements that feed the LMIPDA-VSIMM algorithm for data association and filtering. Monte Carlo simulations demonstrate the effectiveness of the proposed system in an urban scenario contaminated by dense and uneven clutter, strong multipath, and limited line-of-sight.
\end{abstract}

\keywords{Detection, filtering, scheduling, sensing system, signal processing, tracking.} 

\maketitle


\section{INTRODUCTION}
\label{sec:intro}

Tracking airborne targets using sensors has been a well-researched field. Historically, the early efforts began before the advent of World War II and gained precedence in the actual period of war. The field has evolved to incorporate new advancements, and the technology has matured in the ensuing decades. While lot of work has been done in tracking airborne targets, tracking ground targets has drawn sustained attention in the past two decades. Moreover, tracking ground targets in urban terrain poses a new set of challenges and remains relatively unexplored. Target mobility is constrained by road networks, and the quality of measurements is affected by dense and uneven clutter, strong multipath, and limited line-of-sight. In addition, targets can perform evasive maneuvers or undergo a \emph{track swap} owing to congested environments.

Traditionally, the approach to tracking systems design has been to treat sensing and tracking sub-systems as two completely separate entities. While many of the problems involved in the design of such sub-systems have been individually examined in the literature from a theoretical point-of-view, very little attention has been devoted to the challenges involved in the design of an active sensing platform that simultaneously addresses detection, signal processing, tracking, and scheduling in an integrated fashion.

\begin{figure}[hbtp]
\begin{center}
\includegraphics[width=\columnwidth]{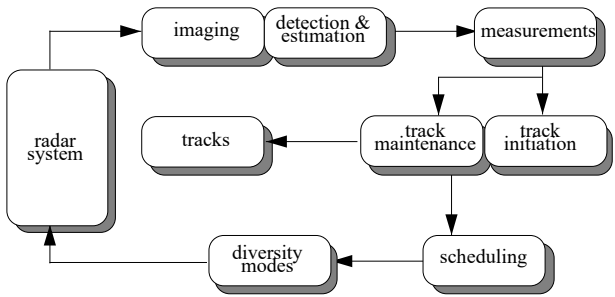}
\caption{Systems-level architecture of the proposed closed-loop active sensing platform.}\label{figloop}
\end{center}
\end{figure}

In this work, we intend to fill this gap by proposing a closed-loop active sensing system for the urban terrain that integrates multitarget detection and tracking, multistatic radar signal processing, and adaptive waveform scheduling with a \emph{one-step-lookahead}. The proposed system simultaneously exploits three distinct levels of diversity: (1) spatial diversity through the use of coordinated multistatic radars; (2) waveform diversity by adaptively scheduling the transmitted radar waveform according to the urban scene conditions; and (3) motion model diversity by using a bank of parallel filters, each one matched to a different motion model. Specifically, at each radar scan, the waveform that yields the minimum trace of the one-step-ahead error covariance matrix is transmitted (termed as one-step-lookahead); the received signal goes through a matched-filter, and curve fitting is used to extract measurements that feed the LMIPDA-VSIMM algorithm for data-association and filtering. The overall system is depicted in Fig.~\ref{figloop}. This feedback structure is fundamentally different from the conventional designs where processing is done sequentially without any feedback.

The primary motivation behind this work is the urban battlefield presented to military forces; however, multiple applications in the transportation, communications, and radiolocation industries (e.g., urban vehicular sensing platforms) would also benefit from effective solutions to the problem of target tracking in urban terrain. The interdisciplinary nature of this work highlights the challenges involved in designing a closed-loop active sensing platform for next-generation tracking and surveillance systems. Further, our work also highlights the importance of incorporating different diversity modes under unfavorable environmental conditions in such platforms.

The remainder of this work is organized as follows. Section~\ref{seclit} presents a discussion on the state-of-the art and past research on active sensing systems for target tracking. In Section~\ref{secmodeling}, we state our modeling assumptions. Section~\ref{secloop} describes the various building blocks of the proposed closed-loop active sensing platform. Simulation results showing the improvements achieved by the proposed system, over a traditional open-loop system that schedules waveform in a round-robin fashion without diversity, are presented in Section~\ref{secsimulation}. Section~\ref{secconc} concludes this work and presents a brief overview of the future work.

\section{LITERATURE REVIEW}
\label{seclit}

In the design of the proposed system, we are faced with multiple challenges in integration of detection, signal processing, tracking, and scheduling. Very few studies have investigated the design of an urban surveillance system that addresses such integration from a systems engineering point-of-view. The literature is populated with studies that address few of the components in this work; nonetheless, a cohesive, comprehensive framework, incorporating all the features described in this study, is missing.

Studies that specifically address closed-loop target tracking are gaining renewed interest from the research community \cite{movpath,rad1,rad2}. Sanders-Reed \cite{sanders2004multitarget} examined integration issues of a multitarget tracking system using video sensors and sensor-pointing commands to close the feedback loop. A more recent work by O'Rourke and Swindlehurst \cite{Rourke} demonstrated a closed feedback loop approach for multitarget tracking with the use of RF/EO sensors. Computational aspects of a closed-loop image-based tracker of airborne targets were considered by Robinson and Sasaki \cite{robinson2000software}, and a closed-loop CCD-based target tracking mechanism for airborne targets was developed by Deng et al. \cite{ccd2017}.  An overview of systems-level modeling for the performance evaluation of closed-loop tracking systems for naval applications was given by Beeton and Hall \cite{BHall}; closed-loop target tracking for underwater systems is also beginning to gain traction \cite{underwat}. Even though all these studies employ some form of tracking, detection, or signal processing, none of them address adaptive waveform scheduling.

Next-generation multifunctional and waveform-agile radars demand innovative resource management techniques to achieve a common sensing goal while satisfying resource constraints. Resource allocation for target tracking has been studied from different perspectives. Adaptive sensing for single target tracking was previously considered by He and Chong \cite{heandchong}, where the sensor scheduling problem was formulated as a partially observable Markov decision process (POMDP). Multitarget results using the same solution framework were presented by Li et al. \cite{liandchong}. Kershaw and Evans \cite{Evans} investigated the problem of one-step-ahead waveform scheduling for tracking systems, while the multi-step-ahead case was considered by Suvorova et al. \cite{Sofia}. The problem of airborne tracking of ground targets was further studied by Miller et al. \cite{Numerica}, where a POMDP framework was used in the coordinated guidance of autonomous unmanned aerial vehicles (UAVs) for multitarget tracking. This work was further expanded by Ragi and Chong \cite{ragiuav,ragidec} and Ragi et al. \cite{ragiaav} to accommodate autonomous amphibious vehicles (AAVs), track swap avoidance, target evasion, and threat-motion model for UAVs; the more general framework of decentralised POMDP (Dec-POMDP) was also studied in one of these research works. A method to incorporate information received from human intelligence assets into the UAV decision making for target tracking was demonstrated by Sarkale and Chong \cite{sarkale} using the POMDP framework.  Allocation of other sensor resources such as target revisit interval and radar steering angle were examined by Kirubarajan et al. \cite{kirbu}.

A case study of urban operations for counter-terrorism, which was analyzed using a probability of attack integrated into a multitarget tracking system, was proposed by Sathyan et al. \cite{terrorism}. Guerci and Baranoski \cite{Baranoski} provided an overview of a knowledge-aided airborne adaptive radar system for tracking ground targets in an urban environment. A lookahead sensor scheduling approach was presented but, as the authors acknowledged, they were ``merely scratching the surface" of a possible solution. Research studies focusing on target tracking in the urban environment have been on the rise ever since. Multipath exploitation for enhanced target tracking in urban terrain was studied by Chakrabortay et al. \cite{chak1}. In the following year, Chakrabortay et al. \cite{chak2} studied target tracking in urban terrain in presence of high clutter. As opposed to the single target tracking cases considered in the previous two works, an integrated method to exploit multipath and mitigate high clutter simultaneously for multitarget tracking was done by Zhou et al. \cite{zhou}. A LIDAR-based approach to track targets at urban intersections was demonstrated recently by Chen et al. \cite{chen18}, where simulation results are provided for real urban intersections. The results in the study suggests that substantial amount of work remains to be done in this field. A data fusion framework for UAVs tracking a single target in an urban terrain was demonstrated by Ramirez-Paredes et al. \cite{data}. Machine learning techniques to track multiple targets in an urban setting have been making a steady progress. Many new ideas in this domain are currently being pursued \cite{machine}.

As is amply justified by the previous studies, a holistic approach that manages all the crucial components involved in target tracking (in urban terrain) is the need of the hour. The only other work that is close in spirit to our work in addressing this problem from a systems perspective is the study by Nielsen and Goodman \cite{complete}, where they combine signal processing, detection, and waveform scheduling for tracking a target. However, our approach significantly differs from the method in that research work. Specifically, we expand upon our previous work \cite{mypaper} and analyze the performance of the proposed closed-loop system under more realistic urban conditions. Ground targets move with enough speed to cause non-negligible Doppler shift; thus, a time-delay versus Doppler image is used to extract range and range-rate measurements. In addition, we analyze the effect of competition among motion models with the inclusion of an acceleration model in the filter design. Waveforms are scheduled based on the trace of the one-step-ahead error covariance matrix, which is proportional to the perimeter of the rectangular region enclosing the covariance ellipsoid. This results in a better approximation of the mean square error than the previously considered matrix determinant. We also add up-sweep and down-sweep chirped waveforms of different pulse durations to the waveform library available for scheduling.

\section{PROBLEM ASSUMPTIONS AND MODELING}
\label{secmodeling}

The first step in designing an active sensing platform for target tracking in an urban environment is to model the various elements that are part of this environment. Even though this process is intrinsically imperfect owing to the many simplifying assumptions explained in this section, the models considered facilitate the analysis of the interplay among these different elements, and how they ultimately affect the overall tracking system performance.

\subsection{Clutter and Multipath in Urban Terrain}
\label{securban}

The overwhelming complexity of the urban environment makes it virtually impossible to consider every detail in every possible scenario. In this work, we consider a representative scenario that allows the tracker to experience the main technical challenges observed in practice: multipath ambiguities, lack of continuous target visibility, and measurement-to-track uncertainty due to clutter.

Throughout this work, the term clutter is used to describe the signal received as a result of scattering from background objects other than targets of interest. Usually dense and unevenly distributed over the surveillance area, urban clutter increases the false alarm rate and missed detections when modeled inappropriately. As opposed to noise, clutter is caused by the transmitted signal; therefore, it is directly related to the signal reflected by targets. We consider clutter as a superposition of $N_{c}$ independent scatterers,
\begin{equation*}
n_{c}(t) = \sum_{i=1}^{N_{c}}a_{i}s\left(t-\tau_{i}\right)e^{2\pi
j\nu_{i}},
\end{equation*}
\noindent where the $i$th scatterer has reflectivity $a_{i}$, $\tau_{i}$ is the time-delay from the transmitter to the $i$th scatterer and back to the receiver, $\nu_{i}$ is the Doppler shift incurred during propagation, and $s(t)$ is the transmitted signal.

Target detection in urban terrain is affected by multipath propagation because of the inability of sensors to distinguish between the received signal scattered directly from a target and the received signal that traversed an indirect path in the urban scenario. However, multipath can be exploited to increase radar coverage and visibility when a direct path between the sensor and target is not available \cite{chak1}. Targets can also be detected due to reflections from buildings, vegetation, and other clutter scatterers having different reflectivity coefficients, which presents different multipath conditions. A high clutter can inflict severe loss on the SNR of the multipath returns \cite{chak2}. Therefore, addressing high clutter and exploiting multipath in an integrated fashion is crucial for urban target tracking.

Using prior knowledge of the terrain, a physical scattering model can be derived. We assume reflective surfaces are smooth, reflectivity coefficients are constant, and the angle of incidence equals the angle of reflection. We further assume that the strength of the radar return is negligible after three reflections.

For an unobstructed target, the direct path can be described as follows. Let $\mathbf{p}$ and $\mathbf{q}$ be vectors corresponding to the paths from transmitter to target and from target to receiver, respectively. The length of the direct path is the sum of the lengths of $\mathbf{p}$ and $\mathbf{q}$; azimuth is the angle between the receiver and $\mathbf{q}$; and the Doppler shift is the sum of the projected target velocity onto $\mathbf{p}$ and $\mathbf{q}$. In all other cases, path length, azimuth, and Doppler shift can be calculated once the reflection point on the clutter scatterer has been determined. For instance, let $\left(x_{c},y_{c}\right)$ be the incidence point of the transmitted signal on the clutter scatterer, $\left(x_{k},y_{k}\right)$ the target position at time step $k$, and $\left(x_{r},y_{r}\right)$ the receiver location. Using simple geometry and the line equation, the reflection point $\left(x_{c},y_{c}\right)$ can be found solving the equations below:
\begin{equation*}\label{eqmultipath1}
y_{c} = mx_{c} + c
\end{equation*}
\begin{equation*}\label{eqmultipath2}
\dfrac{\left[-m(x_{r}-x_{c}) + y_{r}-y_{c}\right]^{2}}{(x_{r}-x_{c})^{2} + (y_{r}-y_{c})^{2}} = \dfrac{\left[-m(x_{k}-x_{c}) + y_{k}-y_{c}\right]^{2}}{(x_{k}-x_{c})^{2} + (y_{k}-y_{c})^{2}},
\end{equation*}
\noindent where $m$ is the slope and $c$ is the $y$-intercept in the line equation representing the clutter scatterer. Both $m$ and $c$ are assumed to be known. Note that such a point may not exist due to possible obscuration and the finite dimensions of scatterers. Equations above refer to the transmitter-target-clutter-receiver path, and the approach is analogous for other paths.

\subsection{Signals for Active Sensing in Urban Terrain}
\label{secsensor}

Radar has become an essential sensor in tracking and surveillance systems in urban terrain owing to its ability to survey wide areas rapidly under any weather conditions \cite{chavali}. In this work, we consider a sensing system where small low-power multistatic radars are distributed over the surveillance area. In particular, we consider bistatic radar pairs augmented by additional sensors (transmitters or receivers). The physical separation between the transmitter and receiver in such a system provides the spatial diversity needed to improve coverage. A bigger coverage area results in an improved detection.

Before we describe the transmitted and received signals, it is important to understand the different time frames involved. While the transmitter and receiver perform signal processing on a intrapulse time frame, the tracker works on a interpulse time frame. Therefore, in the proposed signal model, three time scales are used:  the state sampling period $T = t_{k}-t_{k-1}$, the pulse repetition interval $T_{1}$, and the receiver sampling period $T_{2}$. In general, $T_{2}\ll T_{1}\ll T$. In addition,  we use the far-field assumption and consider the signal wave to be planar.

At time $t_{k}$, a series of pulses is transmitted at periods of $T_{1}$ seconds. Assuming Gaussian-windowed up-sweep and down-sweep chirp signals of unit energy, the signal transmitted by the $n$th transmitter $(n = 1,\ldots,N)$ at time $t_{k}$ is given by:
\begin{equation}\label{txsignal}
s_{k,n}(t) = \sum_{b =
-(B-1)/2}^{(B-1)/2}\frac{\exp\left\{\left[\pm j\gamma -
1/(2\kappa^{2})\right]\left(t-bT_{1}\right)^{2}\right\}}{\left(\pi\kappa^{2}B^{2}\right)^{1/4}},
\end{equation}
\noindent where $t\in\mathbb{R}$, $B$ is the number of pulses transmitted, $\kappa$ represents the pulse duration, and $\gamma$ is the chirp rate. The complex exponential is positive or negative according to the waveform scheduled for transmission: up-sweep chirp or down-sweep chirp, respectively.

The $m$th receiver $(m = 1,\ldots,M)$ is a uniform linear array of $L_{m}$ sensor elements; the sensor elements $L_{m}$ at each receiver are separated by distance $d_{m}$, where the direction of arrival of the signal sent by the $n$th transmitter is $\theta_{n,m}$. We assume coherent processing, i.e., radar returns that arrive at different receiver sampling intervals can be processed jointly. In other words, we are assuming that radar returns can be stored, aligned, and subsequently fed to the receiver for fusion.

The received signal is a summation of reflections from targets of interest and clutter scatterers. Let $P_{k,n,m}$ be the total number of reflections received by the $m$th receiver at time $t_{k}$ that originated from the $n$th transmitter. Signals from the $p$th path $(p = 1,\ldots,P_{k,n,m})$ are subject to a random phase shift $\phi_{n,m}^{p}$ that are uniformly distributed in $(-\pi,\pi]$. Hence, the signal received by the $l$th element $(l = 1,\ldots,L_{m})$ of the $m$th sensor array at time $t_{k} + uT_{2}$ can be written as:
\begin{equation}\label{rxsignal}
\mathbf{y}_{k,m,l}(u) = \sum_{n=1}^{N}\sum_{p =
1}^{P_{k,n,m}}e^{j\phi_{n,m}^{p}}\mathbf{g}_{n,m}^{p}\left(\mathbf{x}_{k};uT_{2}\right)
+\mathbf{e}(u),
\end{equation}
\noindent where $u = (0,\ldots,U-1)$ is the sample index, and $\mathbf{e}(u)$ is a complex white Gaussian process. We can write $\mathbf{g}_{n,m}^{p}\left(\mathbf{x}_{k};uT_{2}\right)$ as:
\begin{eqnarray}\label{eqng}
\lefteqn{\mathbf{g}_{n,m}^{p}\left(\mathbf{x}_{k};uT_{2}\right)=}\\ \nonumber
& &\alpha_{n,m}^{p}\left(\mathbf{x}_{k}\right)\cdot s_{k,n}\left(uT_{2} - \tau_{n,m}^{p}\left(\mathbf{x}_{k}\right)\right)\cdot \\ \nonumber
& &e^{2\pi
j\nu_{n,m}^{p}\left(\mathbf{x}_{k}\right)uT_{2}-j(l_{m}-1)\bar{d}_{m}\left[\cos(\theta_{n,m}^{p}\left(\mathbf{x}_{k}\right))\right]} \cdot\\ \nonumber
& &e^{-2\pi
j\nu_{n,m}^{p}\left(\mathbf{x}_{k}\right)uT_{2}+j(l_{m}-1)\bar{d}_{m}\left[\sin(\theta_{n,m}^{p}\left(\mathbf{x}_{k}\right))\dot{\theta}_{n,m}^{p}\left(\mathbf{x}_{k}\right)uT_{2}\right]},
\end{eqnarray}
\noindent where $\bar{d}_{m} = d_{m}/\lambda$ for the carrier signal wavelength $\lambda$, and $s_{k,n}\left(uT_{2} -
\tau_{n,m}^{p}\left(\mathbf{x}_{k}\right)\right)$ is the delayed replica of the transmitted signal given in \eqref{txsignal}. In addition, the following received signal parameters are defined for the $p$th path between the $n$th transmitter and $m$th receiver: $\alpha_{n,m}^{p}\left(\mathbf{x}_{k}\right)$ is the magnitude of the radar return, including transmitted signal strength and path attenuation; $\tau_{n,m}^{p}\left(\mathbf{x}_{k}\right)$ is the time-delay incurred during propagation; $\nu_{n,m}^{p}\left(\mathbf{x}_{k}\right)$ represents the Doppler shift; and $\theta_{n,m}^{p}\left(\mathbf{x}_{k}\right)$ is the direction of arrival, where $\dot{\theta}_{n,m}^{p}\left(\mathbf{x}_{k}\right)$ is its rate of change. The parameters above can be computed for each target state $\mathbf{x}_{k}$, given prior knowledge of the urban scenario.

\subsection{Models for Target Motion in Urban Terrain}
\label{secmotion}

Target motion in urban terrain can be described by a large number of models that can be combined in various ways. It is not the objective of this work to design novel motion models for targets in urban terrain. Instead, we adopt existing models in the literature. For a comprehensive survey emphasizing the underlying ideas and assumptions of such models, we refer the reader to Li and Jilkov \cite{jilkov}.

Let the target state vector at time $t_{k}$ be
\begin{equation*}
\mathbf{x}_{k} = \left[x_{k},\dot{x}_{k},y_{k},\dot{y}_{k},\ddot{x}_{k},\ddot{y}_{k}\right]^{\top},
\end{equation*}
\noindent where $\top$ denotes matrix transpose, $\left[\dot{x}_{k},\dot{y}_{k}\right]^{\top}$ is the velocity vector, and $\left[\ddot{x}_{k},\ddot{y}_{k}\right]^{\top}$ is the acceleration vector.

Motion models can be divided into two categories: uniform motion (or non-maneuvering) models and maneuvering models. The most commonly used non-maneuvering motion model is the nearly constant velocity (NCV) model, which can be written as:
\begin{equation}\label{eqnstate1}
  \mathbf{x}_{k+1} = \mathbf{F}\mathbf{x}_{k} + \mathbf{G}\mathbf{w}_{k},
\end{equation}
\noindent where the process noise $\mathbf{w}_{k}$ is a zero-mean white-noise sequence,
\begin{equation*}
\mathbf{F} = \left(%
\begin{array}{cccccc}
  1 & T & 0 & 0 & 0 & 0\\
  0 & 1 & 0 & 0 & 0 & 0\\
  0 & 0 & 1 & T & 0 & 0\\
  0 & 0 & 0 & 1 & 0 & 0\\
  0 & 0 & 0 & 0 & 0 & 0\\
  0 & 0 & 0 & 0 & 0 & 0
\end{array}%
\right), \quad \mathbf{G} = \left(%
\begin{array}{cc}
  \frac{1}{2}T^{2} & 0\\
  T & 0\\
  0 & \frac{1}{2}T^{2}\\
  0 & T\\
  0 & 0\\
  0 & 0
\end{array}%
\right),
\end{equation*}
\noindent and $T$ is the state sampling period. The process noise covariance multiplied by the gain is the design parameter
\begin{equation*}
\mathbf{Q} = \mathrm{diag}\left[\sigma^{2}_{w_{x}}\mathbf{Q}^{\prime},\sigma^{2}_{w_{y}}\mathbf{Q}^{\prime}\right],
\end{equation*}
where $\mathbf{Q}^{\prime} = \mathbf{G}\mathbf{G}^{\top}$, and $\sigma^{2}_{w_{x}}$ and $\sigma^{2}_{w_{y}}$ are uncorrelated variances in $x$ and $y$ directions, respectively, corresponding to noisy ``accelerations" that account for modeling errors. To achieve nearly constant velocity or uniform motion, changes in velocity over the sampling interval need to be small compared to the actual velocity, i.e., $\sigma^{2}_{w_{x}}T \ll \dot{x}_{k}$ and $\sigma^{2}_{w_{y}}T \ll~\dot{y}_{k}$.

We consider two different models to describe accelerations and turns. Left and right turns are modeled by the coordinated turn (CT) model with known turn rate $\omega$. This model assumes that the targets move with nearly constant velocity and nearly constant angular turn rate. Although ground target turns are not exactly coordinated turns, the CT model, originally designed for airborne targets, is a reasonable and sufficient approximation for our purposes. Knowledge of each turn rate is based on the prior information about the urban scenario.  For the six-dimensional state vector, the CT model follows \eqref{eqnstate1}, where $\mathbf{w}_{k}$ is a zero-mean additive white Gaussian noise (AWGN) that models small trajectory perturbations, and
\begin{equation*}
\mathbf{F} = \left(%
\begin{array}{cccccc}
  1 & \frac{\sin\omega T}{\omega} & 0 & -\frac{1-\cos\omega T}{\omega} & 0 & 0\\
  0 & \cos\omega T & 0 & -\sin\omega T & 0 & 0\\
  0 & \frac{1-\cos\omega T}{\omega} & 1 & \frac{\sin\omega T}{\omega} & 0 & 0\\
  0 & \sin\omega T & 0 & \cos\omega T & 0 & 0\\
  0 & 0 & 0 & 0 & 0 & 0\\
  0 & 0 & 0 & 0 & 0 & 0
\end{array}%
\right).
\end{equation*}
\noindent As in the NCV model, $\mathbf{Q} =
\sigma^{2}_{w}\mathrm{diag}\left[\mathbf{Q}^{\prime},\mathbf{Q}^{\prime}\right]$, $\mathbf{Q}^{\prime} = \mathbf{G}\mathbf{G}^{\top}$, and $\sigma^{2}_{w}$ is the process noise variance. However, contrary to the NCV model, $x$ and $y$ directions are now coupled.

Accelerations and decelerations are described by the Wiener-sequence acceleration model, where
\begin{equation*}
\mathbf{F} = \left(%
\begin{array}{cccccc}
  1 & T & 0 & 0 & \frac{1}{2}T^{2} & 0\\
  0 & 1 & 0 & 0 & T & 0\\
  0 & 0 & 1 & T & 0 & \frac{1}{2}T^{2}\\
  0 & 0 & 0 & 1 & 0 & T\\
  0 & 0 & 0 & 0 & 1 & 0\\
  0 & 0 & 0 & 0 & 0 & 1
\end{array}%
\right)
\end{equation*}
\noindent and
\begin{equation*}
\mathbf{G} = \left(%
\begin{array}{cc}
  \frac{1}{2}T^{2} & 0\\
  T & 0\\
  0 & \frac{1}{2}T^{2}\\
  0 & T\\
  1 & 0\\
  0 & 1
\end{array}%
\right).
\end{equation*}
\noindent For this model, the process noise $\mathbf{w}_{k}$ in \eqref{eqnstate1} is a zero-mean white-noise sequence with uncorrelated variances in the $x$ and $y$ directions. We consider $\sigma^{2}_{w_{x}}T \ll \ddot{x}_{k}$ and $\sigma^{2}_{w_{y}}T \ll~\ddot{y}_{k}$; under these assumptions, the Wiener-sequence acceleration model is also known as the nearly constant acceleration (NCA) model.

\section{CLOSED-LOOP ACTIVE SENSING PLATFORM}
\label{secloop}

In this section, we outline the main aspects of each component of our closed-loop active sensing platform depicted in Fig.~\ref{figloop}.

\subsection{From Signal Detection to Discrete Measurements}\label{secdetection}

Under the modeling assumption of AWGN, the optimal signal detection is the correlator receiver, or equivalently, the matched-filter \cite{Proakis}. The signal received by the $l$th element of the $m$th sensor array, given in \eqref{rxsignal}, is compared to a template signal by computing a correlation sum of sampled signals. The template signal is a time-shifted, time-reversed, conjugate, and scaled replica of the signal transmitted by the $n$th transmitter at time $t_{k}$:
\begin{equation}\label{replica}
h_{k,n}(t) = as_{k,n}^{*}\left(t_{d}-t\right),
\end{equation}
\noindent where $t_{d}$ is the time-delay incurred during propagation, $^{*}$ represents complex conjugate, and $a$ is the scaling factor assumed to be unity.

In the general multistatic setting with $N$ transmitters and $M$ receivers, we consider the case where the $m$th receiver is a uniform linear array of $L_{m}$ sensor elements. Therefore, we need to combine the signals received by each of the sensor array elements to obtain the total signal received by the $m$th receiver at time $t_{k}$, which can be written as:
\begin{equation}
\mathbf{y}_{k,m} =\sum_{u=0}^{U-1}\sum_{t=1}^{T_{1}}\sum_{l=1}^{L_{m}}\mathbf{y}_{k,m,l}(u)h_{k,n}\left(uT_{2}-t\right),
\end{equation}
\noindent where $\mathbf{y}_{k,m,l}$ is given by \eqref{rxsignal}, and $h_{k,n}$ is given by \eqref{replica}.

Before range and range-rate measurements that feed the tracker can be extracted, pre-processing the radar intensity image is necessary. Assuming each target is a point object (as opposed to an extended object with spatial shape), we use peak detection to locate a point source corresponding to a received power peak on a time-delay versus Doppler image. Because of the strong local (but lack of global) similarities exhibited by urban clutter, image processing techniques aimed to suppress this type of clutter should be based on segmentation analysis, where each image segment must be processed individually in order to distinguish between targets of interest and background scatterers \cite{segmentation}. However, this could be extremely computationally intensive; therefore, we use a more standard form of clutter suppression. Specifically, we calculate a background model prior to tracking using the average of radar intensity images over time to approximate the true urban scenario. The average background image is then subtracted from each image formed using radar returns during the tracking process. For each time-delay $\tau$ and Doppler shift $\nu$, the average magnitude of the radar return is given by:
\begin{equation*}
\bar{A}(\tau,\nu) = \frac{1}{J}\sum_{j=1}^{J}A_{j}(\tau,\nu),
\end{equation*}
\noindent where $j$ indexes times during which the urban scenario was under surveillance prior to tracking, and
\begin{equation*}
A_{k}(\tau,\nu) = A_{k}^{-}(\tau,\nu) - \bar{A}(\tau,\nu)
\end{equation*}
\noindent is the magnitude of the radar return for time-delay $\tau$ and Doppler shift $\nu$ at time step $t_{k}$ during tracking. Note that it is also possible to reduce the image noise, using several image processing techniques, if further improvements on the contrast between background and targets of interest are needed.

Peak detection is implemented iteratively. For each peak $\left(\tau_{k}^{peak},\nu_{k}^{peak}\right)$ found in $A_{k}(\tau,\nu)$, a nonlinear optimization algorithm is used to find a curve that fits the underlying image within a window centered at the peak. When performing curve fitting, we are interested in estimating the measurement error covariance matrix. Since noise sources are assumed to be AWGN, Gaussian curve fitting has been widely used in target detection \cite{gauss}. Note that a Gaussian can be approximated by a quadratic locally within a window centered at the peak. In this work, we fit a two-dimensional quadratic function to each peak in the underlying image. Specifically, at every time step $t_{k}$, we solve the following optimization problem:
\begin{equation*}
\min_{\sigma_{\tau},\sigma_{\nu}}\sum_{\tau}\sum_{\nu}\bigm\vert f_{\sigma_\tau,\sigma_\nu}(\tau,\nu)-A_{k}(\tau,\nu)\bigm\vert^{2},
\end{equation*}
\noindent where for $\epsilon > 0$, the window containing time-delay and Doppler values is defined by $\tau~\in~\left(\tau_{k}^{peak}-\epsilon,\tau_{k}^{peak}+\epsilon\right)$ and $\nu~\in~\left(\nu_{k}^{peak}-\epsilon,\nu_{k}^{peak}+\epsilon\right)$, and
\begin{equation*}
f_{\sigma_\tau,\sigma_\nu}(\tau,\nu) = \sigma_{\tau}^{2}\tau^{2} + 2\sigma_{\tau}\sigma_{\nu}\tau\nu + \sigma_{\nu}^{2}\nu^{2}.
\end{equation*}

We define a scan as the set of measurements generated by a radar receiver from an individual look over the entire surveillance area. The $k$th scan by the $m$th receiver corresponding to its $k$th look is denoted by:
\begin{equation*}
Z_{k,m} =
\left\{\mathbf{z}_{k,m}^{1},\mathbf{z}_{k,m}^{2},\ldots,\mathbf{z}_{k,m}^{N_{k,m}}\right\},
\end{equation*}
where $N_{k,m}$ is the total number of measurements in scan $Z_{k,m}$. In this work, the $j$th $\left(j=1,\ldots,N_{k,m}\right)$ measurement in the $k$th scan of the $m$th receiver is the following two-dimensional vector of range and range-rate:
\begin{equation*}
\mathbf{z}_{k,m}^{j} = \left[
\begin{array}{c}
  r_{k,m}^{j} \\
  \dot{r}_{k,m}^{j}
\end{array}
\right],
\end{equation*}
\noindent where $\left(r_{k,m}^{j},\dot{r}_{k,m}^{j}\right)$ corresponds to the location of the $j$th peak in the time-delay and Doppler image from receiver $m$ by trivial transformation. Associated with each measurement vector $\mathbf{z}_{k,m}^{j}$ is an error covariance matrix
\begin{equation*}
\mathbf{R}_{k,m}^{j}(\psi) = \left[
\begin{array}{cc}
  \sigma_{r_{k,m}^{j}}^{2} & \rho_{r\dot{r}}\left(\sigma_{r_{k,m}^{j}}\sigma_{\dot{r}_{k,m}^{j}}\right)\\
  \rho_{r\dot{r}}\left(\sigma_{r_{k,m}^{j}}\sigma_{\dot{r}_{k,m}^{j}}\right) & \sigma_{\dot{r}_{k,m}^{j}}^{2}
\end{array}
\right],
\end{equation*}
\noindent where $\psi$ is the vector of parameters that characterize the waveform transmitted at $t_{k}$, and $\rho_{r\dot{r}}$ is the correlation coefficient between range and range-rate measurement errors. The vector $\psi$ is included in the measurement noise covariance matrix description to show the explicit dependence of this matrix on the transmitted waveform. In particular, transmitted waveforms defined by \eqref{txsignal} are characterized by pulse duration $\kappa$ and chirp rate $\gamma$; hence, in this case,
\begin{equation*}
\psi = \left[\begin{array}{c}
               \kappa \\
               \gamma
             \end{array}
\right].
\end{equation*}
\noindent The correlation coefficient between measurement errors $\rho_{r\dot{r}}$ also depends on the transmitted waveform and can be calculated using the waveform's ambiguity function \cite{mediterranean}. In particular, the correlation coefficient for the up-sweep and down-sweep chirp waveforms, considered in this work, are strongly negative and positive, respectively.

In state estimation, the measurement model describing the relationship between the target state at time $t_{k}$ and the $k$th radar scan can be written as:
\begin{equation*}
\mathbf{z}_{k,m}^{j} = \mathbf{H}\left(\mathbf{x}_{k}\right) + \mathbf{v}_{k},
\end{equation*}
\noindent where $\mathbf{H}$ is a vector-valued function that maps the target state $\mathbf{x}_{k}$ to its range and range-rate, and $\mathbf{v}_{k}$ is the zero-mean Gaussian measurement noise vector with covariance matrix $\mathbf{R}_{k,m}^{j}(\psi)$.

\subsection{Multitarget-Multisensor Tracker}

We consider a tracker implemented as a sequential filter that weighs measurements in each scan. In addition, tracks are initiated, maintained, and terminated in an integrated fashion. Note that we use the word \textit{track} instead of \textit{target} because we have no a priori knowledge of the number of targets in the urban scenario. Also, algorithms discussed in this section have been previously presented in the literature. Hence, rather than deriving each algorithm below, we highlight their main features related to the design of a closed-loop active sensing system for urban terrain.

\subsubsection{Automatic Track Initiation and Termination}

The goal in track initiation is to estimate tentative tracks from raw measurements without any prior information about how many targets are present in the surveillance area.

We follow the two-point differencing algorithm, according to which it takes two time steps (or two radar scans) for a track to be initiated \cite{Barbook}. For each receiver $m$ $(m = 1,\ldots,M)$, a tentative track is initiated for every declared detection, i.e., for every peak in the time-delay versus Doppler image exceeding a given detection threshold and that cannot be associated with an existing track. In particular, at $t_{1}$ a tentative track $\mathbf{x}_{1}^{(j)}$ is initiated for each measurement $j = 1,\ldots, N_{1,m}$ in scan $Z_{1,m}$. Assuming the velocity of a target along the $x$ and $y$ coordinates lies within the intervals $\left[-\dot{x}_{k-1}^{max},\dot{x}_{k-1}^{max}\right]$ and $\left[-\dot{y}_{k-1}^{max},\dot{y}_{k-1}^{max}\right]$, respectively, a track is initiated at $\left(x_{k},y_{k}\right)$ when
\begin{equation*}
x_{k}\in\left[\left(-\dot{x}_{k-1}^{max}-2\sigma_{\dot{x}_{k-1}}\right)T,\left(\dot{x}_{k-1}^{max}+2\sigma_{\dot{x}_{k-1}}\right)T\right]
\end{equation*}
\noindent and
\begin{equation*}
y_{k}\in\left[\left(-\dot{y}_{k-1}^{max}-2\sigma_{\dot{y}_{k-1}}\right)T,\left(\dot{y}_{k-1}^{max}+2\sigma_{\dot{y}_{k-1}}\right)T\right],
\end{equation*}
\noindent where $T=t_{k}-t_{k-1}$ is the state sampling period, and $\sigma_{\dot{x}_{k-1}}$ and $\sigma_{\dot{y}_{k-1}}$ are the standard deviations of target velocities in $x$ and $y$ directions, respectively. Each measurement that falls into the track initiation area yields an initial position and velocity from which a track is initiated. Measurements can then be associated with this new track starting at $t_{k}>2$, and the target's acceleration can then be estimated at the filtering stage of the tracker, described in Section~\ref{secfiltering}.

The usual approach to track termination is to declare a track terminated if such a track has not been associated with any new measurements for two consecutive time steps. We adopt a more integrated approach and use a probability of track existence, defined in Section~\ref{secdatassoc}, that is initialized for every initiated track. Specifically, a track is terminated if the probability of track existence falls below a given track termination threshold.

\subsubsection{Measurement Validation}

For each initiated track  $\mathbf{x}_{k}^{(t)}$, $t=1,\ldots,\mathrm{T}_{k}$, we define a gate in the measurement space within which measurements to be associated with track $\mathbf{x}_{k}^{(t)}$ are expected to lie. Only those measurements that lie within the gate are said to be validated; therefore, only such measurements are associated to track $\mathbf{x}_{k}^{(t)}$. The size and shape of the gate can be defined in several different ways. We use the ellipsoidal validation gating \cite{ellipsoid} and apply the following statistical test:
\begin{equation*}
\left[\mathbf{z}_{k}^{i}-\hat{\mathbf{z}}_{k}^{(t)}\right]^{\top}\left(\mathbf{S}_{k}^{(t)}\right)^{-1}\left[\mathbf{z}_{k}^{i}-\hat{\mathbf{z}}_{k}^{(t)}\right]<g^{2},
\end{equation*}
\noindent where $\mathbf{z}_{k}^{i}$ represents the $i$th measurement in the $k$th scan, $\hat{\mathbf{z}}_{k}^{(t)}$ is the predicted measurement for track $\mathbf{x}_{k}^{(t)}$, $\mathbf{S}_{k}^{(t)}$ represents the innovation covariance at scan $k$, and $g$ is a threshold computed from Chi-square distribution tables, such that, if a target is detected, its measurement is validated with gating probability $P_{G}$. The number of degrees of freedom of $g$ is equal to the dimension of the measurement vector. In the two-dimensional case, the area of the validation ellipse is $g^{2}\pi\det(\mathbf{S}_{k}^{(t)})^{1/2}$, where $\det$ is the matrix determinant.

\subsubsection{Filtering}
\label{secfiltering}

The tracking algorithm needs to be adaptive in order to handle a time-varying number of targets and dynamic urban conditions. We show in Section~\ref{secsimulation} that the variable structure interacting multiple model (VS-IMM) estimator is effective under such conditions \cite{VSIMM1,VSIMM2}. The VS-IMM estimator implements a separate filter for each model in its model set, which is determined adaptively according to the underlying terrain conditions. Specifically, at each time step $t_{k}$, the model set is updated to:
\begin{equation*}
\mathcal{M}_{k} =
\left\{\emph{r}_{k}\in\mathcal{M}^{total}\bigm|\mathcal{I},\mathbf{x}_{k-1}^{(t,\emph{r})},\mathbf{P}_{k-1}^{(t,\emph{r})},\emph{r}_{k-1}\in\mathcal{M}_{k-1}\right\},
\end{equation*}
\noindent where  $\mathbf{x}_{k-1}^{(t,\emph{r})}$ and $\mathbf{P}_{k-1}^{(t,\emph{r})}$ are the mean and covariance of track $t$ in the filter matched to model $\emph{r}$ at $t_{k-1}$, $\mathcal{I}$ represents prior information about the urban scenario, and $\mathcal{M}^{total}$ is the set of all possible motion models. Changes in track trajectory are modeled as a Markov chain with transition probabilities given by:
\begin{equation*}
\pi_{ij} = P\left\{\emph{r}_{k} = i|\emph{r}_{k-1}=j\right\},\quad
i,j\in\mathcal{M}^{total}.
\end{equation*}

In this work, we consider the unscented Kalman filter (UKF) algorithm. Initially proposed by Julier and Uhlmann \cite{ukf}, the UKF represents the state distribution by a set of deterministically chosen sample points. Each UKF filter, matched to a different motion model, runs in parallel in the VS-IMM framework. The estimated mean and covariance from each model-matched filter are mixed (Gaussian mixture) before the next filtering time step. The overall output of the VS-IMM estimator is then calculated by probabilistically combining the individual estimates of each filter \cite{kim2004imm}.

\subsubsection{Data Association}
\label{secdatassoc}

We consider the linear multitarget integrated probabilistic data association (LMIPDA) algorithm for data association \cite{suv1,suv2}. An extension of the single-target integrated probabilistic data association \cite{IPDA}, LMIPDA models the notion of track existence as a Markov chain. Let $\chi_{k}$ denote the event that a track exists at $t_{k}$. The a priori probability that a track exists at $t_{k}$ is given by:
\begin{equation*}
\psi_{k|k-1}\triangleq\mathrm{P}\left\{\chi_{k}\biggm|\bigcup_{i=1}^{k-1}Z_{i}\right\},
\end{equation*}
where $Z_{k}$ is the set of measurements from all receivers at time $t_{k}$, i.e., $Z_{k} = \cup_{m=1}^{M}Z_{k,m}$. The evolution of track existence over time satisfies the following equations:
\begin{eqnarray*}
\psi_{k|k-1}& = &p_{11}\psi_{k-1|k-1} + p_{21}\left(1-\psi_{k-1|k-1}\right)\\
1-\psi_{k|k-1}& = &p_{12}\psi_{k-1|k-1} +
p_{22}\left(1-\psi_{k-1|k-1}\right),
\end{eqnarray*}
\noindent where $p_{ij}$, $i,j=1,2$, are the corresponding transition probabilities.

The central ideal behind the LMIPDA algorithm is the conversion of a single-target tracker in clutter into a multitarget tracker in clutter by simply modifying the clutter measurement density according to the predicted measurement density of other tracks. The modified clutter density of track $\mathbf{x}_{k}^{(t)}$ given the $i$th measurement can be written as:
\begin{equation*}
\Omega_{i}^{(t)} = \rho_{i}^{(t)} + \sum_{s=1,s\neq
t}^{T_{k}}p_{i}^{(s)}\frac{P_{i}^{(s)}}{1-P_{i}^{(s)}},
\end{equation*}
\noindent where $\rho_{i}^{(t)}$ is the clutter density in the validation gate of track $\mathbf{x}_{k}^{(t)}$ given the $i$th measurement in the $k$th scan $\mathbf{z}_{k}^{i}$, $P_{i}^{(t)}$ is the a priori probability that $\mathbf{z}_{k}^{i}$ is the true measurement for track $\mathbf{x}_{k}^{(t)}$, i.e.,
\begin{equation*}
P_{i}^{(t)} =
P_{D}P_{G}\psi_{k|k-1}^{(t)}\frac{p_{i}^{(t)}/\rho_{i}^{(t)}}{\sum_{i=1}^{N_{k}^{(t)}}{p_{i}^{(t)}/\rho_{i}^{(t)}}},
\end{equation*}
\noindent where $\psi_{k|k-1}^{(t)}$ is the probability of existence of track $\mathbf{x}_{k}^{(t)}$, $p_{i}^{(t)}$ is the a priori measurement likelihood (Gaussian density), and $N_{k}^{(t)}$ is the total number of measurements associated with track $\mathbf{x}_{k}^{(t)}$ at time $t_{k}$. The probability of track existence is calculated as follows:
\begin{equation*}\label{eqexistence}
\psi_{k|k}^{(t)} = \frac{\left(1-\delta_{k}^{(t)}\right)\psi_{k|k-1}^{(t)}}{1-\delta_{k}^{(t)}\psi_{k|k-1}^{(t)}},
\end{equation*}
\noindent where
\begin{equation*}
\delta_{k}^{(t)} = P_{D}P_{G}\left(1-\sum_{i=1}^{N_{k}^{(t)}}\frac{p_{i}^{(t)}}{\Omega_{i}^{(t)}}\right).
\end{equation*}

For each model $\emph{r}\in\mathcal{M}_{k}$, we define the following probabilities of data association:
\begin{equation*}
\beta_{k,0}^{(t,\emph{r})} = \frac{1-P_{D}P_{G}}{1-\delta_{k}^{(t,\emph{r})}}
\end{equation*}
\noindent for clutter measurements, and for each target measurement $i>0$,
\begin{equation*}
\beta_{k,i}^{(t,\emph{r})} = \frac{1-P_{D}P_{G}p_{i}^{(t,\emph{r})}}{\left(1-\delta_{k}^{(t,\emph{r})}\right)\Omega_{i}^{(t)}},
\end{equation*}
\noindent where $p_{i}^{(t,\emph{r})}$ is the a priori likelihood of measurement $i$ assuming association with track $\mathbf{x}_{k}^{(t)}$ that follows motion model $\emph{r}$, i.e.,
\begin{equation*}
p_{i}^{(t)} = \sum_{\emph{r}\in\mathcal{M}_{k}}p_{i}^{(t,\emph{r})},
\end{equation*}
\noindent and
\begin{equation*}
\delta_{k}^{(t,\emph{r})} = P_{D}P_{G}\left(1-\sum_{i=1}^{N_{k}^{(t)}}\frac{p_{i}^{(t,\emph{r})}}{\Omega_{i}^{(t)}}\right).
\end{equation*}

Finally, the motion model for each track $\mathbf{x}_{k}^{(t)}$ is updated according to the following model probabilities:
\begin{equation*}\label{eqmodelprob}
\mu_{k}^{(t,\emph{r})} = \mu_{k|k-1}^{(t,\emph{r})}\frac{1-\delta_{k}^{(t,\emph{r})}}{1-\delta_{k}^{(t)}},
\end{equation*}
\noindent for $\emph{r}\in\mathcal{M}_{k}$.

\subsection{Waveform Scheduling}\label{secschedule}

Many modern airborne radars have a waveform scheduler implemented. Ideally, the scheduler would use a library of waveforms especially designed to improve detection and the overall tracking performance.

We consider the general waveform selection problem, which in the multitarget tracking case can be written as:
\begin{equation*}
\min_{\psi\in\Psi}\frac{1}{\mathrm{T}_{k}}\sum_{t=1}^{\mathrm{T}_{k}}\mathrm{E}\left\{\| \mathbf{x}_{k}^{(t)} - \mathbf{\hat{x}}_{k}^{(t)} \|^{2} \bigm| Z_{k} \right\},
\end{equation*}
where $\Psi$ represents the waveform library, $\mathbf{\hat{x}}_{k}$ is the tracking state estimate, and $\mathrm{T}_{k}$ is the total number of tracks at $t_{k}$.

In particular, we consider the one-step ahead (or myopic) waveform scheduling problem, where the waveform selected for transmission at $t_{k+1}$ is given by:
\begin{equation}\label{wavesch}
\psi_{k+1} = \underset{\psi_{k+1}\in\Psi}{\operatorname{argmin}}~\frac{1}{\mathrm{T}_{k}}\sum_{t=1}^{\mathrm{T}_{k}}\mathrm{Tr}\left\{\mathbf{P}_{k+1}^{(t)}\left(\psi_{k+1}\right)\right\},
\end{equation}
\noindent where $\mathrm{Tr}$ is the matrix trace and $\mathbf{P}_{k+1}^{(t)}$ is the posterior state error covariance matrix corresponding to track $\mathbf{x}_{k}^{(t)}$. For a detailed discussion on the advantages of incorporating lookahead in sensing, see Miller et al. \cite{Numerica}. The performance measure in \eqref{wavesch} is equivalent to minimizing the mean square tracking error over all existing tracks. The posterior state covariance error matrix $\mathbf{P}_{k+1}^{(t)}$ defines a six-dimensional ellipsoid centered at $\mathbf{x}_{k}^{(t)}$ that is a contour of constant probability of error \cite{VanTrees}, and its trace is proportional to the perimeter of the rectangular region enclosing this ellipsoid.

In order to evaluate \eqref{wavesch}, we first approximate the measurement error covariance matrix by the Fisher information matrix $\mathbf{J}(\psi)$ corresponding to the measurement using waveform $\psi$ \cite{Evans,VanTrees}. Specifically,
\begin{equation*}
\mathbf{R}(\psi) = \mathbf{U}\mathbf{J}(\psi)^{-1}\mathbf{U}^{\top},
\end{equation*}
\noindent where $\mathbf{U}$ is the transformation matrix between the time-delay and Doppler measured by the receiver and the target's range and range-rate. In particular, for the up-sweep Gaussian chirp with pulse duration $\kappa$, chirp rate $\gamma$, and wavelength $\lambda$ defined by \eqref{txsignal}, we have:
\begin{equation*}
\mathbf{R}(\psi) = \frac{1}{\eta}\left[\begin{array}{cc}
                     \frac{c^{2}\kappa^{2}}{2} & -\frac{2\pi c^{2}\gamma\kappa^{2}}{\lambda} \\
                     -\frac{2\pi c^{2}\gamma\kappa^{2}}{\lambda} & \left(\frac{2\pi c}{\lambda}\right)^{2}\left(\frac{1}{2\kappa^{2}}+2\gamma^{2}\kappa^{2}\right)
                   \end{array}\right],
\end{equation*}
\noindent where $\eta$ is the signal-to-noise ratio (SNR). A similar expression can be obtained for the down-sweep Gaussian chirp.

The posterior state error covariance matrix can then be calculated for each waveform $\psi\in\Psi$ using the UKF's covariance update equations.

Coherent signal processing across spatially distributed transmitters and receivers is an appealing feature of our sensing platform. The transmission of phase synchronization information for coherent signal processing and waveform scheduling is now possible owing to the advancements in the wireless communications technology, which might have not been possible just a decade before.

\section{SIMULATION RESULTS}
\label{secsimulation}

Monte Carlo simulations are used to evaluate the effectiveness of the proposed closed-loop system in urban terrain.

A number of terrain factors have major impact on the overall system performance. Different road classes (e.g., highways, arterial roads, residential streets, and alleys) impose different constraints on ground vehicles. In addition, different construction materials (e.g., glass, concrete, brick, and wood) have different reflectivity coefficients; therefore, they have different multipath conditions. Any urban environment encompasses multiple components like bridges, footbridges, fences of various types, and round poles etc. Moreover, the structural design of the buildings can vary. The modeling assumptions and the models described in Section \ref{secmodeling} and Section \ref{secloop} might not be able to incorporate every fine detail in an urban environment; nevertheless, they are general enough to accommodate any generic representative urban scenario with adequate detail. For the simulation purposes, we consider a scenario that is representative of the urban conditions to be likely faced by an active sensing tracking system. E.g., we will use vegetation in the simulations to represent the clutter in the urban scenario.

\begin{figure}[hbtp]
\centering
\includegraphics[width=\columnwidth]{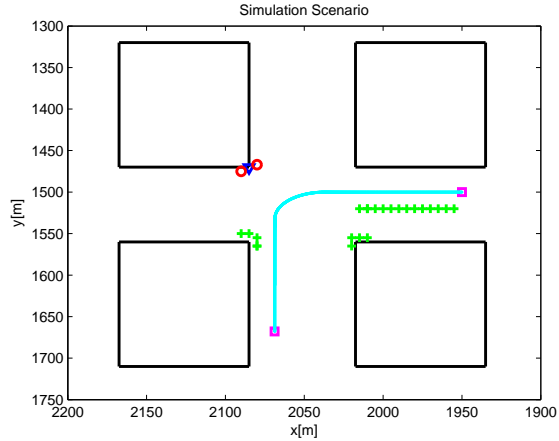}
\caption{The simulated urban terrain. The start and end trajectory points are shown as $\Box$; receivers are shown as $\bigcirc$; the transmitter is shown as $\bigtriangledown$; and clutter discretes are shown as +.}\label{figscenario}
\end{figure}

The simulated scenario is depicted in Fig.~\ref{figscenario}, which shows four building structures at an intersection. The uneven nature of urban clutter is represented by the `+', indicating vegetation on the center median and sidewalks. The overall clutter density is assumed to be 2.5$\text{e}^{-4}\text{m}^{2}$. A radar transmitter, represented by `$\bigtriangledown$', is located at (2085,1470.5); whereas two radar receivers, each with three sensor array elements, are located at (2088,1475) and (2078,1467), both represented by `$\bigcirc$'. The maximum sensor range is 300 meters, and the SNR experienced is 0.2.

Although in reality the transmitted signal can be reflected by multiple scatterers, we assume that the strength of the radar return is negligible after three reflections; therefore, we restrict our simulation to the following paths: transmitter-target-receiver (direct path), transmitter-clutter-receiver, transmitter-target-clutter-receiver, transmitter-clutter-target-receiver, transmitter-clutter-clutter-receiver, transmitter-clutter-target-clutter-receiver, transmitter-target-clutter-clutter-receiver, and transmitter-clutter-clutter-target-receiver.

The simulation experiment consisted of 100 runs, each with a total of 140 radar scans, where a radar scan takes 0.25 seconds. Two targets, 10 seconds apart from each other, are simulated using the same trajectory as follows. Starting at (1950,1500), each target moves at constant velocity of 10 m/s in the $x$ direction for 10 seconds; as they approach the intersection, they start decelerating at constant rate of 1 m/$\text{s}^{2}$ for 5 seconds; they enter a left turn with constant turn rate of $\pi/20$ rad/s for 10 seconds; after completing the turn, each target accelerates for 5 seconds at 1 m/$\text{s}^{2}$ rate; finally, they end their trajectories with constant velocity at (2068.8,1667.8).

Two model sets are used during the motion model adaptation. In the vicinity of intersections, a set consisting of NCA, left-turn CT, and right-turn CT is used. Both the left and right turn CT model are assumed to have a turn rate of $\pi/20$ rad/s. This model set is used between scans 20 and 100. During the remaining radar scans, a set containing the NCA and NCV motion models is used instead. The motion model transition probability matrix is given by
\begin{equation*}
\left(
        \begin{array}{cccc}
          0.99 & 0.01 & 0 & 0 \\
          0.1 & 0.7 & 0.1 & 0.1 \\
          0 & 0.1 & 0.99 & 0 \\
          0 & 0.1 & 0 & 0.99
        \end{array}
      \right).
\end{equation*}

A waveform library consisting of four different Gaussian-windowed chirp signals is considered. Waveforms vary in pulse duration $\kappa$. In particular, radar sensors considered in the simulation experiment are assumed to support the following pulse durations: $\kappa = 0.5 $ $\mu$s, and $\kappa = 1.375$ $\mu$s. In general, longer pulses return more power; however, finer details may be lost. In addition, waveforms of each pulse duration can be either an up-sweep or down-sweep chirp. Pulses are repeated at every 10 milliseconds, and waveforms operate at 4 GHz with 40 MHz of bandwidth.

A more traditional open-loop system, which does not support any diversity modes and that schedules the waveforms in a round-robin fashion, is used as a baseline for comparison. In the baseline implementation, a single UKF using the NCV motion model is considered.

Note that the simulation parameters used in this work do not represent any particular system, and were chosen exclusively for illustration purposes.

\begin{figure}[hbtp]
\centering
\includegraphics[width=\columnwidth]{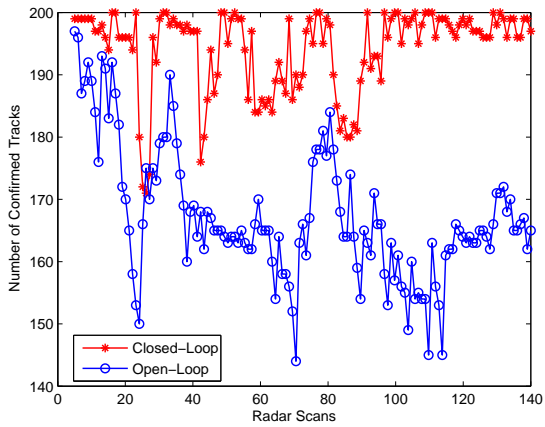}
\caption{Number of confirmed tracks for close-loop and open-loop systems.}\label{figconfirmed}
\end{figure}

\begin{figure}[hbtp]
\centering
\includegraphics[width=\columnwidth]{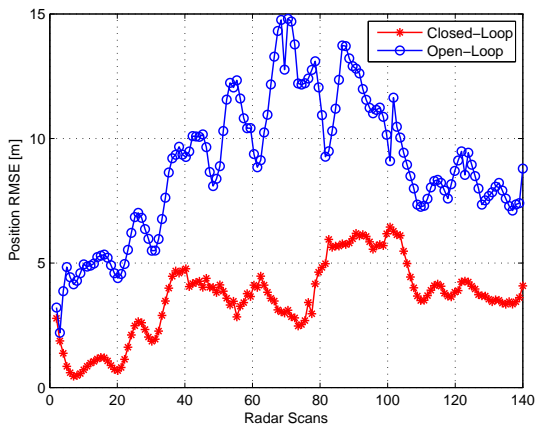}
\caption{Position RMSE for closed-loop and open-loop systems.}\label{figrmse}
\end{figure}

\begin{figure}[hbtp]
\centering
\includegraphics[width=\columnwidth]{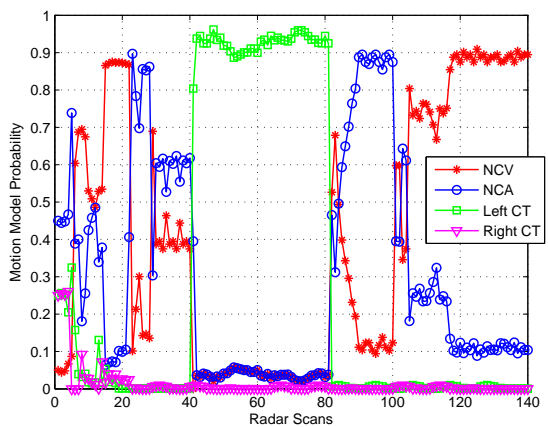}
\caption{Motion model probabilities in the closed-loop system.}\label{figmodels}
\end{figure}

Simulation results in Fig.~\ref{figconfirmed} and Fig.~\ref{figrmse} show that the closed-loop system clearly outperforms its open-loop counterpart. The average number of confirmed tracks is increased by approximately 15\% over 140 radar scans, and the position RMSE is reduced by approximately 60\%. Fig.~\ref{figmodels} shows the evolution of each motion model probability over time. Although there is some ``model competition" between NCV and NCA, the closed-loop system satisfactorily identifies the correct motion model throughout the simulation.

One would like to ascertain the improvement shown by the closed-loop sensing platform over the open-loop system in terms of a particular mode of diversity incorporated in the closed-loop system. Even if this could be sometimes possible by performing multiple simulations under different representative scenarios, in real-world urban environment, assessing the contribution of every single component in the system is impractical. Instead of analysing the contribution of each diversity model separately, we argue that the improvement shown by our system is an emergent feature of the proposed sensing platform, and assessing and fine-tuning the characteristic of each diversity model individually is not required. A theoretical investigation on this is being actively pursued, and we are preparing a separate study to address the same.

\section{Concluding Remarks}\label{secconc}

The closed-loop active sensing system proposed in this work highlights the major challenges in the design of multisensor-multitarget tracking systems, while significantly outperforming its open-loop counterpart. New capabilities for tracking and surveillance, and the seamless integration of different sensing platforms will only be possible with advances in the areas of multisensor data fusion, intelligent algorithms for signal processing and resource allocation, and creative ways to unravel multipath propagation. This work is a first step towards understanding how these research areas interact from a systems engineering perspective to ultimately be integrated into a active sensing tracking platform that operates effectively in urban terrain.

Future work includes tracking dismounts, and the investigation of coordinated non-myopic waveform scheduling schemes for distributed transmitters, which do not have to simultaneously transmit the same waveform. We also consider the expansion of the simulated waveform library, and the formulation of the scheduling problem as a POMDP. Mobile sensing platforms, such as UAVs and vehicular sensor networks, are especially important in hostile urban environments because their completely distributed and opportunistic nature makes it difficult for hostiles to disable surveillance, while potentially increasing the coverage area. Lastly, the performance of the proposed sensing system with the above enhancements will be evaluated through simulations on a real-world urban intersection.

\bibliographystyle{IEEEtran}
\bibliography{IEEEabrv,mypaperbibli}

\end{document}